\documentclass{JHEP3} 
\def\nn{\nonumber}

\def\la{\langle}
\def\ra{\rangle}

\def\l{\left}
\def\r{\right}

\def\beq{\begin{equation}}
\def\eeq{\end{equation}}
\def\bea{\begin{eqnarray}}
\def\eea{\end{eqnarray}}
\def\barr{\begin{array}}
\def\earr{\end{array}}
\def\be{\begin{equation}}
\def\ee{\end{equation}}
\def\bc{\begin{center}}
\def\ec{\end{center}}
\def\dd{\displaystyle}
\def\ov{\overline}

\def\Ne#1{{\cal N}=#1}
\def\De#1{{\it D}=#1}

\def\ZdZd{Z_2\times Z_2}

\def\de{\partial}

\def\Re{\mathop{\mbox{Re}}}

\title{
\boldmath{$\Ne1$} effective potential from dual type--IIA \\ 
D6/O6 orientifolds with general fluxes}
\author{Giovanni Villadoro and Fabio Zwirner \\ 
Dipartimento di Fisica, Universit\`a 
di Roma ``La Sapienza'', and \\ 
INFN, Sezione di Roma, P.le A. Moro 2, I-00185 Rome, Italy \\
E-mail: \email{giovanni.villadoro@roma1.infn.it}, 
\email{fabio.zwirner@roma1.infn.it}}
\preprint{ROMA-1402/05} 
\abstract{ We consider $\Ne1$ compactifications of the type--IIA
theory on the $T^6/(Z_2 \times Z_2)$ orbifold and O6 orientifold, in
the presence of D6--branes and general NSNS, RR and Scherk--Schwarz
geometrical fluxes. Introducing a suitable dual formulation of the
theory, we derive and solve the Bianchi identities, and show how
certain combinations of fluxes can relax the constraints on D6--brane
configurations coming from the cancellation of RR tadpoles. We then
compute, via generalized dimensional reduction, the $\Ne1$, $\De4$
effective potential for the seven main moduli, and comment on the
relation with truncated $\Ne4$ gaugings. As a byproduct, we obtain a
general geometrical expression for the superpotential. We finally
identify a family of fluxes, compatible with all Bianchi identities,
that perturbatively stabilize all seven moduli in supersymmetric
$AdS_4$.}
\begin{document} 
\section{Introduction}
Recently, there has been a lot of interest in orientifold
compactifications of type--II theories, with exact or spontaneously
broken $\Ne1$ supersymmetry, in the presence of branes and fluxes (for
reviews and references, see e.g. \cite{IIrev}). The reason is
twofold. On the one hand, new possibilities appear for constructing
semi--realistic models, with the Standard Model fields emerging from
fluctuations of intersecting and/or magnetized branes. On the other
hand, the possibility of additional RR fluxes with respect to the
heterotic theory provides new opportunities for perturbatively
addressing unsolved problems such as vacuum selection, supersymmetry
breaking and moduli stabilization. The ultimate goal of such a program
would be to obtain explicit, consistent constructions of flux vacua
with broken supersymmetry, an approximately flat four--dimensional
space and all moduli stabilized, with D--branes reproducing the gauge
structure and chiral matter of the Standard Model. We are still far
from this goal, but some first encouraging steps have been made.

In heterotic compactifications (for a review and references, see e.g.
\cite{pol}), and in the known IIB compactifications \cite{IIrev}, the
available perturbative fluxes in the closed string sector are not
sufficient to produce $\Ne1$ supersymmetric $AdS_4$ vacua and to fix
all the geometrical moduli. On the other hand, for type--IIA
compactifications examples exist of stable $AdS_4$ vacua, obtained via
perturbative fluxes only \cite{bc,dkpz,lustsi}.

The authors of \cite{bc} considered IIA compactifications on suitable
manifolds with non-trivial torsion and $SU(3)$ or $SU(2)$ structures,
to get flux vacua with unbroken $\Ne1$ supersymmetry. In particular,
they found an $AdS_4$ supersymmetric vacuum on a nearly--K\"ahler
manifold when all NSNS and RR fluxes are turned on. Further
investigations along similar lines were carried out in \cite{lustsi}.

The authors of \cite{dkpz} focused on the D6/O6 orientifold with the
$T^6/(Z_2 \times Z_2)$ orbifold, which combines a simple geometry of
the internal space with a very rich structure of fluxes. Among the
possibilities allowed by the general system of NSNS, RR and
geometrical Scherk--Schwarz fluxes, they got a stable $AdS_4$
supersymmetric vacuum with all seven geometrical moduli fixed, as well
as non-supersymmetric vacua corresponding to no--scale models or
runaway potentials.

O6 compactifications on the $T^6/(Z_2 \times Z_2)$ orbifold have also
a phenomenological interest, since semi--realistic chiral $\Ne1$ models
with intersecting D6--branes have been formulated in such a framework
\cite{csu}. It is then interesting to investigate the relation between
the compactified $\De10$ theory and the resulting effective $\De4$
theory in more detail: this is the goal of the present paper. For
simplicity, we will set to zero all the brane excitations and the
associated magnetic fluxes: such a setup is obviously insufficient to
give a realistic model, but is a good training ground for a
quantitative and self--consistent discussion of vacuum selection,
supersymmetry breaking and moduli stabilization.

Since our work complements and extends the work of \cite{dkpz}, we
first summarize the results of the latter. That paper focused on the
O6 orientifold of the type--IIA theory (which would give rise, on the
six--torus $T^6$, to an effective $N=4$, $D=4$ supergravity),
consistently combined with the $T^6/(Z_2 \times Z_2)$ orbifold, to
preserve an exact or spontaneously broken $\Ne1$, $\De4$
supergravity. First, it classified the most general fluxes in the
closed string sector that are invariant under the combined orbifold
and orientifold projections: they include fluxes of the NSNS 3--form
$H$; fluxes of the even RR forms $G^{(0)}$ (corresponding to the mass
parameter of massive IIA supergravity \cite{romans}), $G^{(2)}$,
$G^{(4)}$, $G^{(6)}$; also, geometrical fluxes $\omega$, associated
with the internal components of the spin connection, and corresponding
to Scherk--Schwarz \cite{ss} or, equivalently, twisted tori
compactifications. Then, it focused the attention (as we do in the
present paper, leaving further generalizations to future work) on the
effective theory for the seven main moduli coming from the closed
string sector. In particular, it determined the general form of the effective
K\"ahler potential and superpotential, after identifying a suitable
field basis. Instead of considering the Bianchi identities and the RR
tadpole cancellation conditions of the compactified $\De10$ theory, it
explored the constraints on the $\Ne1$ superpotential corresponding to
consistent gaugings of the underlying $\Ne4$, $\De4$ effective
supergravity. For toroidal heterotic compactifications, this was
proved to be a successful strategy, since the Bianchi identities of
the reduced theory, in the presence of general fluxes, are exactly
equivalent \cite{km} to the conditions for a $\Ne4$ gauging. This is
also true in several M-theory and type--II examples, as recently
discussed in \cite{dagfer}.

The goal of the present paper is to consider the same class of IIA
compactifications as in \cite{dkpz}, investigating the structure of
the $\De4$ effective theory for the seven main moduli directly by
generalized $\Ne1$ dimensional reduction, rather than from the point
of view of a truncated gauged $\Ne4$ supergravity. To deal with
general systems of fluxes, D6--branes and O6--planes, we will introduce
a dual formulation of the IIA theory. A similar approach was followed
in \cite{bkorv}, without geometrical fluxes, and in \cite{gh} for
compactifications to two dimensions with RR fluxes only. Our method
will allow us to consider, beyond the limits of \cite{dkpz},
consistent $\Ne1$ compactifications that cannot be viewed as
truncations of $\Ne4$ supergravities. It will also allow us to give a
geometrical interpretation to the effective superpotential, and to
discuss the subtle interplay between the Bianchi identities of the
compactified $\De10$ theory and the conditions for a consistent
gauging of the effective $\Ne4$ theory, in the cases where the $\Ne1$
effective theory can be obtained by a truncation of the former.

Our paper is organized as follows.

In section 2, we define the class of IIA compactifications under
consideration, and, after fixing the conventions and recalling the
standard form of (massive) IIA supergravity, we write down the bulk
and brane actions in the dual version.

In section 3, we derive the Bianchi identities in the presence of
general fluxes (NSNS, RR, geometrical) and of D6/O6 systems compatible
with $\Ne1$ supersymmetry. Their general solution has the form
\bea
H&=&d B+\omega\ B + \ov H\,, \nn\\
G^{(p)} &=& d C^{(p-1)} + \omega \ C^{(p-1)} + H \wedge C^{(p-3)} + 
(\mathbf{\ov G} \ e^{-B})^{(p)} \, , 
\eea
for NSNS and RR fields respectively, with $\omega$, $\ov H$ and 
$\ov G^{(p)}$  constant fluxes subjected to the integrability 
conditions:
\bea
&\omega \ \omega=0\,,\qquad \omega \ \ov H=0\,,&  \nn \\
&\omega\ \ov G^{(p)}+\ov H \wedge \ov G^{(p-2)}=0\, .& 
\eea
D6--brane and O6--plane contributions only show up in the non--trivial
Bianchi identities for the \emph{phantom} components of $G^{(2)}$ that
are projected out by the orbifold:
\beq
\frac12 \l (d G^{(2)}+\omega \, G^{(2)} + 
H G^{(0)} \r)=\sum_a^{D6,O6} \ \mu_a \delta^{(3)} (\pi_a) \, .
\eeq
Their integrability conditions provide additional constraints for
fluxes and localized charges, corresponding to modified RR tadpole
cancellation conditions.

In section 4, we derive the $\Ne1$, $\De4$ effective potential via
generalized dimensional reduction, and check that it agrees with the
general form of the effective K\"ahler potential and superpotential,
obtained in \cite{dkpz} via consistent truncation of an underlying
$\Ne4$, $\De4$ gauged supergravity. Our derivation allows to rewrite
the generic $\Ne1$ effective superpotential in a suggestive
geometrical form, which can be explicitly checked on our example and
generalizes those previously derived (or conjectured on the basis of
duality) by various authors \cite{gh, generalwIIA1, generalwIIA2}:
\beq
W = \frac14 \int_{{\cal M}_6} \ov{\mathbf{G}} \, e^{i J^c}
- i \l ( \ov H -i\omega \, J^c \r) \wedge \Omega^c \, ,
\eeq
where $J^c$ and $\Omega^c$ are the complexified K\"ahler 2--form and 
the holomorphic 3--form, respectively.
We also find that the Bianchi identities, including those for the
geometrical fluxes alone, can be satisfied by systems of fluxes and
branes that do not correspond to $\Ne4$, $\De4$ gaugings, thereby
relaxing some of the constraints on the parameters of the $\Ne1$
superpotential.

In section 5 we produce examples of consistent systems of fluxes,
satisfying all the Bianchi identities (including those for the
geometrical fluxes), that give rise to stable supersymmetric $AdS_4$
vacua with all seven geometrical moduli fixed. In these examples,
those corresponding to $\Ne4$ gaugings involve all types of allowed
fluxes. If we ask only for the Bianchi identities, without requiring
an underlying gauged $\Ne4$ supergravity, then the allowed fluxes that
stabilize all seven main moduli depend on more parameters, and
stabilization can be achieved even if some types of fluxes are
switched off.

In the final section we briefly summarize our results and comment on
the prospects for extending the present work.

%
\section{IIA theory with D6/O6 on \boldmath{$T^6/(Z_2 \times Z_2)$} 
and its dual formulation}
The bosonic sector of $\De10$ type--IIA supergravity consists of the
following fields: in the NSNS sector, the (string-frame) metric
$g_{MN}$, the 2--form potential\footnote{Since there is no risk of
confusion, we do not attach a ``$(2)$'' to $B$ and a ``$(3)$'' to its
field-strength $H$.} $B$ and the dilaton $\Phi$; in the RR sector, the
$(2k+1)$--form potentials $C^{(2k+1)}$, ($k=0,\dots,4$). Here and in
the following, we stick to the conventions of \cite{pol} unless
otherwise stated, and we split the $\De10$ space-time indices as $M =
[ \mu = 0, \ldots , 3\,;\, i = 5, \ldots, 10]$. Actually, the degrees
of freedom of the RR potentials are not all independent, being related
by Poincar\'e duality. The corresponding action can then be expressed
as a function of the independent fields only, usually identified with
$C^{(1)}$ and $C^{(3)}$. This leads to the standard formulation of
$\De10$ IIA supergravity, whose action reads, in a schematic but
self-explanatory notation:
\bea
S_{IIA} & = & S_{NS} + S_{R} + S_{CS} \, , 
\label{eq:standardIIA}
\\ 
S_{NS} & = & \int d^{10}x \, e_{10} \, \frac{e^{-2\Phi}}{2} \, 
\l[ R_{10} + 4 \, (\de \Phi)^2 - \frac12 \, (H)^2 \r] \, ,
\label{eq:Snsns} 
\\
S_{R} & = & - \frac14 \int d^{10}x \, e_{10} \, 
\l[ M^2 + ( F^{(2)})^2 + (\widetilde{F}^{(4)})^2 \r] \, ,
\label{eq:Srrst}
\\
S_{CS} & = & - \frac14 \int d^{10}x \,
B \, \wedge F^{(4)} \wedge F^{(4)} \, ,
\label{eq:Scsst} 
\eea
where
\bea 
H & = & d B \, , \nn 
\\ 
F^{(2)} & = & d C^{(1)} + M \, B \, , \nn 
\\ 
F^{(4)} & = & d C^{(3)} + \frac12 \, 
M \, B \wedge B \, , \nn 
\\ 
\widetilde F^{(4)} & = & F^{(4)} - C^{(1)} \wedge H \, , 
\eea
and $M \equiv F^{(0)}$ is the mass parameter of massive IIA
supergravity \cite{romans}.

In the presence of localized objects, the standard formulation of
(massive) IIA supergravity is not the optimal choice, since
D$(p-1)$--branes carry RR charges and naturally (i.e. electrically)
couple to $p$--form potentials. Thus, depending on the type of
D$(p-1)$--branes considered in the effective theory, it is more
convenient to use a suitably dualized action that includes these
fields, along the lines of ref.~\cite{bkorv}.

The general procedure for the construction of the dual action works as
follows. The first step consists in choosing the independent degrees
of freedom: for each couple of $p$--form potentials $C^{(p)}$
($p=n,8-n$), a single $p$--form potential and its dual field strength
$G^{(9-p)}$ have to be chosen. Both these fields are treated as
independent, with the potential entering the action with the role of a
Lagrange multiplier. The second step is to construct the action which
reproduces the standard formulation of (\ref{eq:standardIIA}) in the
``no--flux'' limit, after elimination of the Lagrange multipliers. The
equations of motion for the RR potentials actually give the Bianchi
identities (BI in the following) for their dual field strengths. The
solutions of these BI give then the explicit expressions for the
$(9-p)$--form field strengths, in terms of a $(8-p)$--form potential
and a combination of fluxes and other fields.

In this work we are interested in the representative case of type--IIA
supergravity with O6--planes and D6--branes, compactified on the
orbifold $T^{6}/(\ZdZd)$ in the presence of general fluxes. However,
as will be clear in the following, most of our results have more
general validity.

We start by defining the action of the orbifold projection on the
internal coordinates $x^i$ ($i = 5, \dots, 10$) of the factorized
six--torus $T^6 = T^2 \times T^2 \times T^2$ as
\beq
Z_2 \ : \;\; (z^1,z^2,z^3) \to (-z^1,-z^2,z^3) \, , 
\qquad
Z'_2 \ : \;\; (z^1,z^2,z^3) \to (z^1,-z^2,-z^3) \, , 
\eeq
where
\beq
z^1 = x^5 + i \ x^6 \, , 
\qquad 
z^2 = x^7 + i \ x^8 \, , 
\qquad 
z^3 = x^9 + i \  x^{10} \, ,
\eeq
and the action of the orientifold involution as
\beq
{\cal R} \ : \;\; (z^1,z^2,z^3) \to  
(- \ov z^1, - \ov z^2, - \ov z^3) \, .
\eeq
The minus sign in the above equation has been chosen to match the
conventions of \cite{dkpz}.  The action of ${\cal R}$ is compatible
with the presence of O6--planes with internal coordinates
$(6,8,10)$. The action of the orbifold group produces extra O6--planes
in the three directions [$(6,7,9)$, $(5,8,9)$, $(5,7,10)$]. The
intrinsic parities of the fields with respect to the orientifold
projection are: $+1$ for the $g_{MN}$, $\Phi$ and the RR $p$--form
potentials with $p=3,7$; $-1$ for $B$ and the RR $p$--form
potentials with $p=1,5,9$.

The bulk scalar fields surviving the orbifold and orientifold
projections are: the dilaton $\Phi$; three K\"ahler moduli $t_A$
($A=1,2,3$) and three complex structure moduli $u_A$ from the diagonal
components $g_{ii}$ ($i=5,\ldots,10$) of the internal metric; three
axions $\tau_A$ from the NSNS 2--form potentials\footnote{In our
short--hand notation, $X_{I|\cdot|\cdot|\cdots}$, where $X$ is a
generic field and $I$ is a group of space--time indices, means the
component $X_{I}$ and all the other components with the same number of
indices as $I$ and the same parity under the orbifold and orientifold
projections. For example, $B_{56| \cdot | \cdot} \equiv B_{56}, \,
B_{78} , \, B_{910}$, and $C^{(3)}_{6810|\cdot|\cdot|\cdot} \equiv
C_{6810}, \, C_{679}, \, C_{789}, \, C_{5710}$.} $B_{56| \cdot |
\cdot}$; one plus three axions $\sigma$ and $\nu_A$ from the RR 3--form
potentials $C^{(3)}_{6810| \cdot | \cdot | \cdot}$.

In the case under consideration, the dualization in the RR sector does
not affect the NSNS part of the action, which remains the same as in
(\ref{eq:Snsns}).  For the part of the action involving the RR sector
we adopt a dual formulation along the lines of \cite{bkorv}. As we
will see, this formalism will allow us to consistently derive both the
effective action in the presence of D6--branes and general fluxes
(NSNS, RR, geometrical), and the BI for the RR field strengths.

For our purposes, we need only the action for the components of the
fields that survive the orbifold and orientifold projections. The
complete list of these fields is given in table~\ref{tab:fields},
\TABULAR[ht]{||c|| c | c || c | c | c ||}{
\hline \hline 
&\multicolumn{2}{c||}{}&\multicolumn{3}{c||}{} \\
$C^{(-1)}\leftrightarrow C^{(9)}$ & 
\multicolumn{2}{c||}{$C^{(1)}\leftrightarrow C^{(7)}$} & 
\multicolumn{3}{c||}{$C^{(3)}\leftrightarrow C^{(5)}$} \\ 
&\multicolumn{2}{c||}{}&\multicolumn{3}{c||}{} \\ \hline &&&&& \\
$\times$ & $\times$ & $\times$ & $\times$ & 
$C^{(3)}_{6810|\cdot|\cdot|\cdot}$ & $C^{(3)}_{\mu\nu\rho}$ \\ &&&&& \\
$G^{(0)}$ & $G^{(2)}_{56|\cdot|\cdot}$ & $(\times)$ & 
$G^{(4)}_{5678|\cdot|\cdot}$ & $G^{(4)}_{\mu\{6810|\cdot|\cdot|\cdot\}}$ & 
$G^{(4)}_{\mu\nu\rho\sigma}$ \\ &&&&& \\
$G^{(10)}$ & $G^{(8)}_{\mu\nu\rho\sigma\{5678|\cdot|\cdot\}}$ & 
$\times$ & $G^{(6)}_{\mu\nu\rho\sigma\{56|\cdot|\cdot\}}$ & 
$G^{(6)}_{\mu\nu\rho\{579|\cdot|\cdot|\cdot\}}$ & $G^{(6)}_{5678910}$ 
\\ &&&&& \\
$C^{(9)}_{\mu\nu\rho\{5678910\}}$ & 
$C^{(7)}_{\mu\nu\rho\{5678|\cdot|\cdot\}}$ & 
$C^{(7)}_{\mu\nu\rho\sigma\{6810|\cdot|\cdot|\cdot\}}$ & 
$C^{(5)}_{\mu\nu\rho\{56|\cdot|\cdot\}}$ & 
$C^{(5)}_{\mu\nu\{579|\cdot|\cdot|\cdot\}}$ & $\times$ 
\\ &&&&& \\ \hline \hline
}{\label{tab:fields} \it IIA bulk fields surviving the orbifold and
orientifold projections. Each column may contain the invariant
components of $p$--form potentials, the corresponding $(p+1)$--form field
strengths and their duals. Each dot indicates a different combination
of indices compatible with the projections.}
%
from which we can see that the choice of independent fields is almost
completely fixed by the projections. The only components where a
choice is allowed are in fact $C^{(3)}_{6810|\cdot|\cdot|\cdot}$ and
$C^{(5)}_{\mu\nu\{579|\cdot|\cdot|\cdot\}}$. For these components,
taking as Lagrange multiplier either $C^{(3)}$ or $C^{(5)}$ is
equivalent to arranging the corresponding $D=4$ axionic fields into
either a linear or a chiral multiplet, respectively. We choose the
second option, which makes contact with the standard formulation of
the effective $\Ne1$, $\De4$ supergravity.

Notice also that, at variance with the cases discussed in
\cite{bkorv}, for the dual couple $(C^{(5)},C^{(3)})$ we must choose
different dualizations for different components, consistently with the
action of the orbifold and orientifold projections on the space-time
coordinates.

Finally, there is a subtlety with the components $C^{(7)}_{\mu \nu
\rho \sigma \ \{6810|\cdot|\cdot|\cdot \} }$, whose dual field
strengths $G^{(2)}$ are truncated away by the orientifold and orbifold
projections. Although these fields cannot contribute to the effective
scalar potential, their derivatives can survive the projections and,
as we will show, they do indeed lead to non--trivial BI.

Summarizing, the couples of independent bulk fields that we keep in
the action are
\bea
&&G^{(0)} \leftrightarrow C^{(9)}_{\mu\nu\rho\{5678910\}}\,, \nn \\
&&G^{(2)}_{56|\cdot|\cdot} \leftrightarrow 
C^{(7)}_{\mu\nu\rho\{78910|\cdot|\cdot\}}\,, \qquad
G^{(2)} \leftrightarrow 
C^{(7)}_{\mu\nu\rho\sigma\{6810|\cdot|\cdot|\cdot\}}\,, \nn \\
&&G^{(4)}_{5678|\cdot|\cdot} \leftrightarrow 
C^{(5)}_{\mu\nu\rho\{910|\cdot|\cdot\}}\,, \qquad 
G^{(4)}_{\mu\{6810|\cdot|\cdot|\cdot\}} \leftrightarrow 
C^{(5)}_{\mu\nu\{579|\cdot|\cdot|\cdot\}}\,, \nn \\
&&G^{(6)}_{5678910} \leftrightarrow C^{(3)}_{\mu\nu\rho}\,.
\label{eq:chosen}
\eea

We are now ready to write our RR bulk action. In analogy with 
\cite{bkorv}, we perform the field redefinition
\be
\label{eq:ac}
\mathbf{A} = e^B \ \mathbf{C} \, ,
\ee
which simplifies the calculations. As in \cite{bkorv}, we group
potential and field strengths into formal sums such as $\mathbf{A} =
\sum_{p = odd} A^{(p)}$, $\mathbf{C} = \sum_{p = odd} C^{(p)}$,
$\mathbf{G} = \sum_{p = even} G^{(p)}$. Equations involving these
formal sums must be then considered term by term, projecting onto the
forms of each given order $p$ and considering only the relevant
combinations. Similarly, we use the short--hand notation $e^B = 1 + B
+ (1/2) \, B \wedge B + \ldots$, and in expressions such as $e^B \,
\mathbf{C}$ or $e^B \, \mathbf{G}$ the wedge product is
understood. Then:
\beq
\label{eq:SRR}
S_{R} + S_{CS} \quad  \longrightarrow  \quad S_{kin}(G) + 
S_{CS}(G,B) + S_{LM}(A,G) \, , 
\eeq
\bea
S_{kin}(G) & = & -\frac14 \int d^{10}x \  e_{10}\ 
\l [ (G^{(0)})^2+(G^{(2)})^2+(G^{(4)})^2+(G^{(6)})^2 \r]
\, , \label{eq:Skin} \\ 
S_{LM}(A,G) & = & \frac12 \int \l ( A^{(9)}-A^{(7)}+A^{(5)}-A^{(3)} 
\r )\wedge d ( \mathbf{G} \ e^{B})\,, \label{eq:SLM} 
\eea
where sums restricted to the components in eq.~(\ref{eq:chosen}) are
implicit. In the part of the action containing the Lagrange
multipliers, $S_{LM}(A,G)$, a sum over all the $G^{(p)}$ is
understood, in order to fill the ten--dimensional wedge product.
After the orbifold and orientifold projections, and with the choice of
independent bulk fields specified in eq.~(\ref{eq:chosen}), there is
no surviving term from the ten--dimensional Chern--Simons action
$S_{CS}(G,B)$ in eq.~(\ref{eq:SRR}).

Finally, we can add contributions from D6--branes and O6--planes.
Neglecting D--brane fluctuations and localized fluxes, they read
\beq 
\label{eq:SD6O6}
S_{D6/O6} = S_{DBI} + S_{WZ} =
\sum_{a}^{D6,O6} \ \l [-T_a \int_{\pi_a} d^{7}x 
\  e_7 \ e^{-\Phi} + \mu_a \int_{\pi_a} A^{(7)}\r] \, ,
\eeq
where $T_a=\mu_a$ and branes/planes wrap factorisable 3-cycles
$\pi_a=\prod_{A=1}^3(n_A^a,m_A^a)$.

In summary, in the dual formulation the total action reads
\beq
S^{dual}_{IIA} = S_{NS} + S_{kin}(G) + 
S_{LM}(A,G) +  S_{D6/O6}(A) \, ,
\label{eq:Sdual}
\eeq
with the four different contributions given by eqs.~(\ref{eq:Snsns}),
(\ref{eq:Skin}), (\ref{eq:SLM}) and (\ref{eq:SD6O6}), respectively.
%
\section{Bianchi identities and their solutions}
Before discussing the BI for the RR sector, we first look at the BI
for the NSNS 3--form, which reads:
\beq 
d H=0 \, , 
\eeq
with general solution given by: 
\beq 
H = d B + \ov H \, ,
\qquad
\ov{H} = {\rm constant} \, .
\eeq
In a trivial dimensional reduction, the first contribution gives the
kinetic terms for the axions in $B$, and the second one is a constant
flux. However, for generalized Scherk--Schwarz reductions (or twisted
tori), characterized by constant parameters $\omega_{jk}^{\quad i}$
(called $f^{i}_{\quad kj}$ in \cite{ss} and $C_{ijk}$ in \cite{dkpz}),
the BI is modified as follows:
\beq 
d H + \omega H = 0 \, .
\label{eq:BINSNSw} 
\eeq
In the above equation and in the following, the transition from
trivial to generalized dimensional reduction is described by the
simple rule \cite{ss}:
\bea 
d X&\to& d X+\omega \ X \,,\nn \\ \omega \ X &\equiv&
\omega_{\bullet \, \bullet}^{\quad \times} X_{\times\, \bullet\,
\cdots\, \bullet}\,, 
\eea
where $X$ is a generic form and indices indicated with a full dot are
antisymmetrized, while those with a cross are contracted. The
constants $\omega$ (which may be called {\em geometrical fluxes}) can
thus be viewed as operators that map a $p$--form into a $(p+1)$--form.
According to the discussion of \cite{ss}, appropriate for toroidal
compactifications as we are considering here, the $\omega$ components
must be structure constants of a Lie group, satisfying the Jacobi
identity:
\be
\omega \ \omega \equiv
\omega_{\bullet \, \bullet}^{\quad \times}
\omega_{\times \, \bullet}^{\quad \circ}
= 0 \, ,
\label{eq:jacw}
\ee
where an empty dot denotes a free index, and the additional condition
\be
\omega_{\times \circ}^{\quad \times} = 0
\label{eq:sstrace}
\ee
must hold. While eq.~(\ref{eq:sstrace}) is automatically satisfied by
the geometrical fluxes that survive our orientifold and orbifold
projections, eq.~(\ref{eq:jacw}) imposes a non--trivial constraint. As
discussed in \cite{km}, this constraint can be interpreted as the BI
for the non--trivial connection of the internal manifold, $d^2=0$, for
constant geometrical fluxes $\omega$. The general solution to
eq.~(\ref{eq:BINSNSw}) is:
\beq
H=dB+\omega B + \ov H \, ,
\qquad
\ov{H} = {\rm constant} \, ,
\eeq
provided that:
\beq
\omega \ \ov H=0 \, , 
\label{eq:BIcH}
\eeq
which represents the integrability condition for the BI of $H$. In
the class of compactifications under discussion, the solution for the
surviving components of $H$ reads:
\bea 
(\mu\{56|78|910\})~: & & H  =  d B \, , \nn 
\\ 
(579|5810|6710|689)~: & & H=\omega \ B + \ov H \, ,
\label{eq:BifluxH} 
\eea
and eq.~(\ref{eq:BIcH}) is trivially satisfied because of the
projections. As in \cite{dkpz}, we have four NSNS fluxes
$\ov{H}$ and twelve geometrical fluxes $\omega$,
\bea
&\omega_{68}^{\quad 10} \,, \quad 
\omega_{10 6}^{\quad 8} \,,\quad
\omega_{8 10}^{\quad 6} \,,& \nn \\  
&\omega_{57}^{\quad 10} \,, \quad 
\omega_{95}^{\quad 8} \,, \quad 
\omega_{79}^{\quad 6} \,,& \nn \\
&\omega_{58}^{\quad 9} \,, \quad 
\omega_{89}^{\quad 5} \,, \quad 
\omega_{67}^{\quad 9} \,, \quad 
\omega_{96}^{\quad 7} \,, \quad 
\omega_{105}^{\quad 7} \,, \quad
\omega_{710}^{\quad 5} \,, &
\eea
allowed by the orbifold and orientifold projections. However, as will
be discussed in more detail in the next section, the constraints on
these fluxes coming from the BI are not exactly the same as those
derived in \cite{dkpz} from truncated $\Ne4$ gaugings.

To derive the BI for the RR field strengths $G^{(p)}$ from the dual
action (\ref{eq:Sdual}), it is sufficient to vary it with respect to
the corresponding Lagrange multipliers $A^{(9-p)}$. We can see that,
if $A^{(9-p)}$ does not couple to localized sources, the general BI
for $G^{(p)}$ read:
\bea
&& d (e^B \ \mathbf{G})=0 \nn \\
&\Rightarrow& d G^{(p)} + H \wedge G^{(p-2)}=0 \, ,
\eea
which implies
\beq
d G^{(p)} +\omega \ G^{(p)} + 
H \wedge G^{(p-2)} = 0 \, , 
\label{eq:BIG}
\eeq
for generalized Scherk--Schwarz dimensional reduction. The
general solution to eq.~(\ref{eq:BIG}) is:
\beq
G^{(p)}=d C^{(p-1)} + \omega \ C^{(p-1)} + 
H \wedge C^{(p-3)} + ( \mathbf{\ov G} \ e^{-B})^{(p)} \,,
\quad
(\mathbf{\ov G} = {\rm constant}) \, ,
\label{eq:BIGsol}
\eeq
provided that the following integrability condition holds:
\beq
\omega \ \ov G^{(p)} +\ov H \wedge \ov G^{(p-2)} =0 \,. 
\label{eq:BIcG}
\eeq
The last term in eq.~(\ref{eq:BIGsol}) is understood as expanded and
projected into a $p$--form wedge product. The solution of
eq.~(\ref{eq:BIGsol}) is valid in general, and shows the power of the
dual construction. Each $p$--form field strength in the action
(\ref{eq:Skin}) receives contributions from a combination of all other
fluxes with $p'\le p$ and, as we will see in the next section, these
mixing terms will be crucial to reconstruct the full $\Ne1$
supersymmetric expression of the effective potential. In our case,
the explicit solutions for the surviving components of the field
strengths $G^{(p)}$ read:
\bea 
&& G^{(0)} = \ov G^{(0)} \, , \nn \\
(56|\cdot|\cdot)~: & & 
G^{(2)} = \ov G^{(2)} - B \ \ov G^{(0)} \,, \nn \\
(5678|\cdot|\cdot)~: & & 
G^{(4)} = \omega\,C^{(3)}+\ov G^{(4)}-B\wedge \ov
G^{(2)}+\frac12 B\wedge B \, \ov G^{(0)}\,, \nn \\
(\mu\{6810|\cdot|\cdot|\cdot\})~: & & 
G^{(4)}  = d C^{(3)} \, , \nn \\
(5678910)~: & & 
G^{(6)} = \ov H \wedge C^{(3)}+(\omega\,B)\wedge C^{(3)}+\ov
G^{(6)} \nn\\ &&- B\wedge \ov G^{(4)} +\frac12 B\wedge B\wedge \ov
G^{(2)} -\frac16 B\wedge B\wedge B \, \ov G^{(0)}\,, \label{eq:solG}
\eea 
while the corresponding constraints of eq.~(\ref{eq:BIcG}) are
all trivially satisfied after the projections.

Finally, we need to discuss separately the case of the \emph{phantom}
components of $G^{(2)}$, whose dual Lagrange multipliers $A^{(7)}$
couple to D6--branes and O6--planes. The corresponding BI read:
\bea
\frac12 \l (d G^{(2)}+\omega \, G^{(2)} + H G^{(0)} \r)=
\sum_a^{D6,O6} \ \mu_a \delta^{(3)} (\pi_a) \,, 
\eea
where $\delta^{(3)}(\pi_a)$ is the 3--form Poincar\'e--dual to the
3-cycle $\pi_a$ of the brane/plane $a$. We are not interested in
solving this equation, since the solution for $G^{(2)}$ is odd under
the combined orientifold and orbifold projections and does not enter
the effective potential. However, the integrability conditions
associated to the above BI give, in components, the following
non-trivial constraints:
\bea
\frac12 \l ( \omega \, \ov G^{(2)}+\ov H \, \ov G^{(0)}\r)_{579}
&=&\sum_a^{D6,O6} \mu_a m_1^a m_2^a m_3^a\,, \nn \\
\frac12 \l ( \omega \, \ov G^{(2)}+\ov H \, \ov G^{(0)}\r)_{5810}
&=&\sum_a^{D6,O6} \mu_a m_1^a n_2^a n_3^a\,, \nn \\
\frac12 \l ( \omega \, \ov G^{(2)}+\ov H \, \ov G^{(0)}\r)_{6710}
&=&\sum_a^{D6,O6} \mu_a n_1^a m_2^a n_3^a\,, \nn \\
\frac12 \l ( \omega \, \ov G^{(2)}+\ov H \, \ov G^{(0)}\r)_{689}
&=&\sum_a^{D6,O6} \mu_a n_1^a n_2^a m_3^a\,, \label{eq:BId6o6}
\eea	
where $(n_A^a,m_A^a)$ are the wrapping numbers of the brane/plane $a$
on the $A$--th 2-torus.

These constraints show how the RR-tadpole cancellation conditions for
D6--branes are modified in the presence of general fluxes. It is
interesting to notice that there exist combinations of fluxes that
escape the usual charge cancellation conditions for D6--branes and
O6--planes on each 3-cycle. This is analogous to what happens in
type--IIB compactifications with both RR and NSNS 3--form fluxes
turned on \cite{gkp}. In that case, however, the contribution of
fluxes to both the scalar potential and the BI is either vanishing or
positive-definite, leading to no--scale models or runaway
potentials\footnote{Strictly speaking, this is true for the Calabi-Yau
compactifications considered in \cite{gkp} and for the one considered
here (i.e. $T^6/\ZdZd$), but other possibilities can be studied (for a
recent example, see e.g. \cite{eg}).}. On the other hand, in type--IIA
compactifications the richness of the flux structure allows us to
avoid such bounds, enlarging the possibilities for moduli
stabilization and model building. It would be also interesting to see
whether these modifications of the BI can admit stable supersymmetric
configurations of intersecting D6--branes on toroidal
compactifications without orbifolds.
%
\section{\boldmath{$\Ne1$} effective potential and superpotential}
We can now derive the effective $\De4$ action for the bulk scalar
fields of our theory by substituting the solutions of the BI, as found
in the previous section, into the action of eq.~(\ref{eq:Sdual}). The
kinetic terms are fixed by the K\"ahler potential\footnote{Notice that
our conventions for $Y$ differ from those of \cite{dkpz} by a constant
multiplicative factor $2^7$, which can be reabsorbed as an overall
multiplicative factor in the superpotential $W$.}
\beq
K= - \log \, Y \, , \qquad 
Y= s \, t_1 \, t_2 \, t_3 \, u_1 \, u_2 \, u_3 \, ,
\eeq
where the seven complex scalars
\beq
S=s+i\sigma\,, \qquad T_A=t_A+i \tau_A\,, 
\qquad U_A=u_A+i\nu_A\,, \qquad (A=1,2,3) \, ,
\label{eq:main}
\eeq
are connected to the ten--dimensional fields via:
\beq
g_{MN}= {\rm blockdiag}\l ( \hat s^{-1}\, \widetilde{g}_{\mu\nu},
\ t_1 \,\hat u_1,\ \frac{t_1}{\hat u_1}, \ t_2 \, \hat u_2,\ 
\frac{t_2}{\hat u_2},\ t_3 \, \hat u_3,\ \frac{t_3}{\hat u_3} 
\r) \,, 
\eeq
\bea
e^{-2\Phi} & = & \frac{\hat s}{t_1 t_2 t_3} \,, 
\qquad B_{56|78|910}=\tau_{1|2|3} \,, 
\qquad C^{(3)}_{6810}=\sigma \,, 
\qquad C^{(3)}_{679|589|5710}=-\nu_{1|2|3} 
\,,  \nn \\
s & = & \sqrt{\frac{\hat s}{\hat u_1\,\hat u_2\,\hat u_3}} \,, 
\quad u_1 = \sqrt{\frac{\hat s \ \hat u_2\,\hat u_3}{\hat u_1}} \, , 
\quad u_2 = \sqrt{\frac{\hat s \ \hat u_1\,\hat u_3}{\hat u_2}} \, , 
\quad u_3 = \sqrt{\frac{\hat s \ \hat u_1\,\hat u_2}{\hat u_3}} \, ,
\eea
and $\widetilde{g}_{\mu\nu}$ is the metric in the $\De4$ Einstein
frame. In this paper, we call {\it main} moduli the seven complex
scalars in (\ref{eq:main}), {\it geometrical} moduli their real parts,
{\it axions} their imaginary parts.

The $\De4$ scalar potential receives contributions from the
Einstein term, the generalized kinetic term containing the NSNS three
form, the analogous terms for the RR field strengths, and finally the
tensions of localized objects:
\beq
V = V_E + V_H + V_G + V_6 \,.
\eeq
The first contribution is the well--known Scherk--Schwarz potential
\cite{ss}:
\beq
V_E = \frac18\frac{ e_{10}}{\widetilde{e}_4}e^{-2\Phi}\l 
(\omega_{jk}^{\quad i}\, \omega_{j'k'}^{\ \quad i'} \  
g_{ii'}\   g^{jj'} \  g^{kk'} + 2 \ \omega_{jk}^{\quad i}
\ \omega_{j'i}^{\quad k} \  g^{jj'}\r)\,,
\eeq
which involves only the geometrical fluxes $\omega$. The second 
contribution comes from the $(H)^2$ term in eq.~(\ref{eq:Snsns}):
\beq
V_H = \frac14\frac{ e_{10}}{\widetilde{e}_4}e^{-2\Phi} \l 
(\ov H+\omega\,B\r)^2\,.
\eeq
Then we have the contributions coming from $S_{kin}$ of the RR sector:
\bea
V_G & = & \frac14\frac{ e_{10}}{\widetilde{e}_4}\l [\l (\ov G^{(0)}\r)^2
+\l (\ov G^{(2)} - B \, \ov G^{(0)}\r)^2 +\r. \nn \\
&& + \l (\omega\,C^{(3)}+\ov G^{(4)}-B\wedge \ov G^{(2)}
+\frac12 B\wedge B \, \ov G^{(0)}\r)^2 + \nn \\
&& +\l (\ov H \wedge C^{(3)}+(\omega\,B)\wedge C^{(3)}
+\ov G^{(6)} - B\wedge \ov G^{(4)} + \r. \nn \\
&& \quad \l.\l.  +\frac12 B\wedge B\wedge \ov G^{(2)}
-\frac16 B\wedge B\wedge B \, \ov G^{(0)}\r)^2 \r ] \, .
\eea
Finally, we have the contributions coming from D6--branes and
O6--planes:
\beq
V_{D6/O6}=\sum_a \, T_a \frac{ e_{7}^{(a)}}{\widetilde{e}_4}e^{-\Phi} \, ,
\eeq
where $e_{7}^{(a)}$ refers to the induced metric on the
D6--brane/O6--plane under consideration. The potential $V_{D6/O6}$
coming from localized sources can be conveniently rewritten by
exploiting the requirement that the brane setup preserves $\Ne1$
supersymmetry. In this case, in fact:
\beq
V_{D6/O6}=e^{K} t_1 t_2 t_3 \sum_a T_a \l ( m_1^a m_2^a m_3^a 
\ s - m_1^a n_2^a n_3^a \ u_1 - n_1^a m_2^a n_3^a \ u_2 - 
n_1^a n_2^a m_3^a \ u_3 \r)\,,
\eeq
where, recalling that $\mu_a=T_a$ because of supersymmetry, we find
the same combination of wrapping numbers present in the BI of
eq.~(\ref{eq:BId6o6}). We can then rewrite the localized contributions
to the potential as:
\bea \label{eq:Vd6o6flux}
V_{D6/O6}&=&\frac12 e^K t_1 t_2 t_3 \l [ (\omega \ov G^{(2)}
+\ov H \, \ov G^{(0)})_{579}\ s -  (\omega \ov G^{(2)} +
\ov H \,\ov G^{(0)})_{5810}\ u_1 \r. \nn \\
&& \qquad \qquad \l. - (\omega \ov G^{(2)}
+\ov H \, \ov G^{(0)})_{6710} \ u_2 - 
(\omega \ov G^{(2)}+\ov H \,\ov G^{(0)})_{689} \ u_3 \r]\,, 
\eea
which is \emph{independent} of the particular setup of D6--branes and
O6--planes, as long as they preserve $\Ne1$. Their contribution to the
potential enters only indirectly, via the constraints on the fluxes
spelled out in eq.~(\ref{eq:BId6o6}).  We verified that the terms in
eq.~(\ref{eq:Vd6o6flux}) are precisely those that need to be added to
the bulk potential, $V_E+V_H+V_G$, to reconstruct the standard $\Ne1$
formula:
\beq \label{eq:potw}
V=e^{K}\l [ \sum_{\alpha=1}^7 \l|W_\alpha(\varphi) + 
K_\alpha W(\varphi) \r|^2 - 3 |W(\varphi)|^2\r] \, ,
\qquad
\varphi^{1,\ldots,7} = (S,T_1,T_2,T_3,U_1,U_2,U_3) \, ,
\eeq
where $W_\alpha\equiv \de W / \de \varphi^\alpha$, 
$K_\alpha\equiv \de K / \de \varphi^\alpha$, with superpotential
\bea
4 \, W&=&
\Bigl ( \omega_{810}^{\quad 6} \, T_1 \, U_1+\omega_{106}^{\quad 8} 
\, T_2 \, U_2+\omega_{68}^{\quad 10} \, T_3 \, U_3\Bigr)
- S \, \Bigl (\omega_{79}^{\quad 6} \, T_1 + 
\omega_{95}^{\quad 8} \, T_2 +\omega_{57}^{\quad 10} \, T_3 \Bigr)
 \nn \\
&& -\Bigl ( \omega_{89}^{\quad 5} \, T_1 \, U_3+
\omega_{96}^{\quad 7} \, T_2 \, U_3 + \omega_{710}^{\quad 5} 
\, T_1 \, U_2+\omega_{67}^{\quad 9} \, T_3 \, U_2
+\omega_{105}^{\quad 7} \, T_2 \, U_1+
\omega_{58}^{\quad 9} \, T_3 \, U_1\Bigr)\nn \\
&&
+ i \Bigl ( \ov G^{(4)}_{78910} \, T_1
+\ov G^{(4)}_{91056} \, T_2+\ov G^{(4)}_{5678}\, T_3\Bigr)-
\Bigl (\ov G^{(2)}_{56} \, T_2 \, T_3+\ov G^{(2)}_{78}\,  T_1 \, T_3
+\ov G^{(2)}_{910} \, T_1 \, T_2 \Bigr) \nn \\
&&
+ \ov G^{(6)} - i \, \ov G^{(0)} \, T_1 \, T_2 \, T_3
+i \Bigl ( \ov H_{579} \, S - \ov H_{5810} \, U_1
- \ov H_{6710} \, U_2- \ov H_{689} \, U_3 \Bigr)  \,,
\label{eq:superpot}
\eea
provided that eq.~(\ref{eq:jacw}) is satisfied. We stress the
importance of the mixing terms coming from the solutions of the BI in
eq.~(\ref{eq:solG}) for recovering the entire potential of
eq.~(\ref{eq:potw}) by dimensional reduction.

The general superpotential of eq.~(\ref{eq:superpot}) reduces to the
one found in \cite{dkpz}, under the simplifying assumption of a
symmetry with respect to the interchange of the three factorized
two--tori (plane--interchange--symmetry). It also takes a suggestive
and compact form when rewritten in terms of geometrical objects:
\beq 
\label{eq:superpotgeo}
W = \frac14 \int_{{\cal M}_6} \ov{\mathbf{G}} \, e^{i J^c}
- i \, \l( \ov H -i \omega \, J^c \r) \wedge \Omega^c \, ,
\eeq
where, in our conventions, $J^c$ and $\Omega^c$ read
\bea
J^c & = & J + i \ B \, ,
\qquad
J = {i \over 2} \sum_{A=1}^3 dz^A \wedge d \ov{z}^A \, ,
\nn \\
\Omega^c & = & \Re \left( i \ e^{\dd - \Phi} \ \Omega \right) 
+ i \ C^{(3)} \, , \qquad
\Omega = dz^1 \wedge dz^2 \wedge dz^3 \, .
\eea
In particular, in our case:
\be
J^c_{56|78|910} = T_{1|2|3} \, , 
\ee
\be
\Omega^c_{6810} = S \, , 
\qquad 
\Omega^c_{679|589|5710}= - U_{1|2|3} \, .
\ee
Since $\Re(iJ^c)=-B$, we can recognize in eq.~(\ref{eq:superpotgeo}) the
combinations of fluxes and fields entering the BI of
eqs.~(\ref{eq:BifluxH}) and (\ref{eq:BIGsol}). Indeed:
\beq
W = \frac14 \int_{{\cal M}_6} \widetilde G^{(6)}\,,
\eeq
where $\widetilde G^{(6)}$ is the solution $G^{(6)}$ of the BI 
in eq.~(\ref{eq:solG}), with the substitutions:
\beq
B\to - i J^c \,,\qquad C^{(3)}\to - i \ \Omega^c \,.
\eeq
Moreover, since by T-duality:
\beq
\ov{\mathbf{G}} \ e^{iJ^c} \quad \leftrightarrow \quad
\ov G^{(3)} \wedge \Omega_{IIB} \, ,
\eeq
eq.~(\ref{eq:superpotgeo}) exhibits the same structure of the
Gukov--Vafa--Witten \cite{gvw} IIB superpotential:
\beq
\int_{{\cal M}_6} \l (\ov G^{(3)} - i S \ov H \r)  
\wedge \Omega_{IIB} \, ,
\eeq
apart from the twisted tori contribution proportional to $\omega \,
J^c \wedge \Omega^c =dJ^c \wedge \Omega^c$. Parts of the
superpotential of eq.~(\ref{eq:superpotgeo}) have already been argued
in other contexts by using mirror symmetries or dimensional reduction
with torsion (see e.g. \cite{gh,generalwIIA1,generalwIIA2}). All this
makes us think that the validity of eq.~(\ref{eq:superpotgeo}) extends
beyond the $T^6/(\ZdZd)$ orbifold considered here, although we
explicitly derived it only for this case.

The explicit presence of all the K\"ahler ($J^c$), complex structure
and dilaton ($\Omega^c$) moduli in the IIA superpotential of
eq.~(\ref{eq:superpotgeo}) explains why in this theory it is possible
to find stable vacua, with all geometrical moduli fixed, in the
perturbative regime, and makes type--IIA compactifications appealing
candidates for perturbatively addressing the problem of moduli
stabilization.

\subsection{Relation with $\Ne4$ gaugings}
We can now study the relations between the conditions found in \cite{dkpz}
by requiring that the effective field theory be a truncation of a gauged
$\Ne4$ supergravity, and those found here by dimensional
reduction and BI, requiring only $\Ne1$ supersymmetry.

Ref.~\cite{km} found that in the heterotic case there is a
one--to--one correspondence between Jacobi identities of the $\Ne4$
gauged supergravity and Bianchi identities in the presence of general
fluxes. As we will see in a moment, this is no longer true, in
general, for type--II compactifications. In \cite{dkpz}, the formalism
of gauged $\Ne4$ supergravity was used to derive, upon consistent
truncation, the superpotential of eq.~(\ref{eq:superpot}). That work
also found the consistency conditions on the fluxes implied by the
closure of the Jacobi identities of the gauged $\Ne4$, $\De4$ Lie
algebra, at the compensator level. In \cite{dagfer}, the analysis of
\cite{km} was extended to M--theory and to more general type--II
compactifications. From these analyses, we can conclude that the
constraints coming from $\Ne4$ supergravity can be divided into two
groups. The first group contains some of the equations in
(\ref{eq:jacw}), while the second group is made of the last three BI
of eq.~(\ref{eq:BId6o6}) with vanishing r.h.s. In contrast with the
heterotic case, the conditions for a consistent $\Ne4$ $\De4$ gauging
do not seem sufficient to reconstruct all the consistency conditions
coming from toroidal generalized dimensional reductions, in particular
the closure of all the Scherk--Schwarz BI (\ref{eq:jacw}). It is not
known if these type of $\Ne4$ gaugings can have a geometrical
interpretation. On the other hand, the conditions from the second
group are more restrictive than those found here from generalized
$\Ne1$ dimensional reduction. In this case, however, we can give a
definite interpretation: since different branes/planes preserve
different $\Ne4$ subsets of the $\Ne8$ bulk supersymmetry, depending
on their orientation, the vanishing of the r.h.s. in the last three BI
of eq.~(\ref{eq:BId6o6}) corresponds to requiring that all the
D6--branes preserve the same $\Ne4$. If this condition is not
satisfied, the effective field theory cannot be seen as a simple
truncation of $\Ne4$ gauged supergravity. It is worth noticing,
however, that the $\Ne1$ formalism of \cite{dkpz} is able to reproduce
the correct form of the effective superpotential even for general
brane configurations, at least for the bulk fields. Consistency
conditions for the fluxes, on the other hand, require the derivation
of the BI from reduction, analogously to what has been done in the
present work.

\section{Examples of consistent $\Ne1$ vacua}
In ref.~\cite{dkpz}, a stable supersymmetric $AdS_4$ vacuum
(satisfying the criteria\footnote{Notice that requiring a stable
supersymmetric $AdS_4$ vacuum does not imply that all the
background values of the moduli are fixed.} of \cite{stads4})
was constructed, starting from the superpotential of
eq.~(\ref{eq:superpot}), in the plane--interchange--symmetric
limit. It was found that for the following choice of fluxes:
$$
-\frac19 \ov G^{(6)} = \ov G^{(2)} = \frac16 \omega_1 
= -\frac12 \omega_2 \, ,
$$
$$
\frac13 \ov G^{(4)} = \frac15 \ov G^{(0)}=
-\frac12 \ov H_{0} = \frac12 \ov H_{1} \, ,
$$
\beq
\omega_3=0 \,,
\label{eq:adsdkpz}
\eeq
where, in our notation:
\bea
\omega_1&\equiv&\omega_{68}^{\quad 10}=
\omega_{10 6}^{\quad 8}=\omega_{8 10}^{\quad 6}\,, \nn \\
\omega_2&\equiv&\omega_{57}^{\quad 10}=
\omega_{95}^{\quad 8}=\omega_{79}^{\quad 6}\,, \nn \\
\omega_3&\equiv&\omega_{58}^{\quad 9}=
\omega_{89}^{\quad 5}=\omega_{67}^{\quad 9}=\omega_{96}^{\quad 7}=
\omega_{105}^{\quad 7}=\omega_{710}^{\quad 5}\,, \nn \\
\ov H_0&\equiv&H_{579} \,, \nn \\
\ov H_1&\equiv&H_{5810|6710|689}\,,
\eea
there exists a stable supersymmetric $AdS_4$ vacuum at $\la S\ra=\la
T_A\ra=\la U_A\ra=1$, with $\langle V \rangle =- 3 \langle e^K |W|^2
\rangle $, such that all seven geometrical moduli are fixed. As
explained in \cite{dkpz}, the quantization condition on the fluxes can
be made manifest by a suitable rescaling of the moduli. Moreover, the
fluxes in eq.~(\ref{eq:adsdkpz}) automatically satisfy the $\Ne4$
constraints:
\beq 
\label{eq:ne4constr}
\omega_3 \l ( \omega_3-\omega_1\r)=0\,, 
\qquad  5\, \ov H_1^2 = 3 \,\omega_2^2\, .
\eeq
The first one corresponds to one of the geometrical BI of
eq.~(\ref{eq:jacw}). The second one represents the absence of
localized RR sources from 3--cycles non--parallel to the $\Ne4$
O6--plane (6,8,10). However, for the vacuum solution to be fully
consistent with a Scherk--Schwarz reduction, also the following
constraint coming from the geometrical BI (\ref{eq:jacw}) must be
satisfied:
\beq
\omega_2 \l (\omega_1-\omega_3\r) = 0 \, .
\label{eq:sscond}
\eeq

It is not difficult to find solutions that extend
eq.~(\ref{eq:adsdkpz}) and also satisfy eq.~(\ref{eq:sscond}). If the
fluxes are chosen so that:
\beq
\frac19 \ov G^{(6)} = -t_0^2\,\ov G^{(2)} = \frac{t_0\,u_0}{6} \omega_1 
= \frac{s_0\,t_0}{2} \omega_2= \frac{t_0\,u_0}{6} \omega_3\,,
\label{eq:fluxone}
\eeq
\beq
\frac{t_0}{3} \ov G^{(4)} = \frac{t_0^3}{5} \ov G^{(0)}
=-\frac{s_0}{2} \ov H_{0}=\frac{u_0}{2} \ov H_{1}\,, 
\label{eq:adsss}
\eeq
all the BI are satisfied and there is a supersymmetric $AdS_4$ vacuum
at $\la S\ra=s_0$, $\la T_A\ra=t_0$ and $\la U_A\ra=u_0$
($s_0\,,\,t_0\,,\, u_0\in \mathbb{R}^+$). All seven main moduli are
stabilized in the sense of \cite{stads4}. Moreover, all seven
geometrical moduli are fixed and, for generic values of the two
independent parameters in eqs.~(\ref{eq:fluxone}) and
(\ref{eq:adsss}), also all axions are fixed, with the exception of a
single residual Goldstone mode, corresponding to a combination of
$(\sigma,\nu_1,\nu_2,\nu_3)$. The constraints of
eqs.~(\ref{eq:fluxone}) and (\ref{eq:adsss}) represent a family of
supersymmetric $AdS_4$ solutions of the model compatible with the
Scherk--Schwarz reduction. Other solutions can be obtained by
performing ``electric'' duality transformations such as axionic shifts
or dilatations, and by relaxing the constraints from
plane--interchange--symmetry.

The second relation of eq.~(\ref{eq:ne4constr}) reads now
\beq
5\,u_0^2\,\ov H_1=3\,s_0^2\,t_0^2\,\omega_2^2\,.
\eeq
When it is satisfied, the $AdS_4$ vacua of eq.~(\ref{eq:adsss}) also
admit an interpretation in terms of $\Ne4$ gaugings. However, the
generalized $\Ne1$ reduction performed in the present paper allows us
to study a wider class of solutions that do not admit such an
interpretation.  For example, we could choose either $\ov H_1=0$ or
$\omega_2=0$, and compensate for the mismatch of RR charges with
suitable D6--brane configurations. Summarizing, we can construct
stable $\Ne1$ $AdS_4$ vacua by switching on either RR and geometrical
fluxes ($\ov G^{(6)}$, $\ov G^{(2)}$ and $\omega_{1,2,3}$), or RR and
NSNS fluxes ($\ov G^{(4)}$, $\ov G^{(0)}$ and $\ov H_{0,1}$), or
both. However, for the solution to admit a $\Ne4$ gauging
interpretation, all the possible types of fluxes have to be turned on
simultaneously.

Vacua corresponding to no--scale models or to runaway potentials can
be easily found as well, analogously to what was done in
\cite{dkpz}. Also in these cases, however, there is not a one--to--one
correspondence between the solutions compatible with generalized
$\Ne1$ dimensional reduction and those satisfying the $\Ne4$ gauging
conditions of the $\De4$ effective field theory.

%
\section{Conclusions and outlook}

In this paper we considered $\Ne1$ compactifications of type--IIA
supergravity on the O6 orientifold, consistently combined with the
$T^6/(Z_2 \times Z_2)$ orbifold, in the presence of D6-branes and of
general NSNS, RR and Scherk--Schwarz geometrical fluxes. We introduced
a suitable dual action, to derive and solve the BI, and to compute, by
$\Ne1$ generalized dimensional reduction, the effective scalar
potential and superpotential. We derived an elegant geometrical
expression for the superpotential, which generalizes previous
results. We also identified a family of fluxes that are compatible
with all the BI, including the geometrical ones, and perturbatively
fix all seven geometrical moduli of the closed string sector on a
stable supersymmetric $AdS_4$ vacuum. We finally clarified the
relation of this approach with the one of \cite{dkpz}, based on the
constraints coming from $\Ne4$ gaugings.

Among the possible extensions of the present work, needed to proceed
towards realistic models, the most immediate ones seem to be the
inclusion of brane fields, the inclusion of localized magnetic fluxes
and the calculation of soft terms (along the lines of \cite{magn,
soft}).

Another open question, which may be more difficult to address, is the
extension of our results to compactifications with non--negligible
warping.
%
\acknowledgments
We thank G.~Dall'Agata, J.--P.~Derendinger, S.~Ferrara and C.~Kounnas
for discussions. This work was supported in part by the European
Programme ``The Quest For Unification'', contract MRTN-CT-2004-503369.
%


%

\begin{thebibliography}{99}
%
\bibitem{IIrev}
C.~Angelantonj and A.~Sagnotti, {\it Open strings}, Phys.\ Rept.\ {\bf
371}, 1 (2002) [Erratum-ibid.\ {\bf 376}, 339 (2003)]
[arXiv:hep-th/0204089];
\\ 
A.~M.~Uranga, {\it Chiral four-dimensional string compactifications
with intersecting D--branes}, Class.\ Quant.\ Grav.\ {\bf 20} (2003)
S373 [arXiv:hep-th/0301032];
\\
F.~G.~Marchesano Buznego, {\it Intersecting D-brane models},
arXiv:hep-th/0307252;
\\
A.~R.~Frey, {\it Warped strings: Self-dual flux and contemporary
compactifications}, arXiv:hep-th/0308156;
\\ 
E.~Kiritsis, {\it D--branes in standard model building, gravity and
cosmology}, Fortsch.\ Phys.\ {\bf 52} (2004) 200
[arXiv:hep-th/0310001];
\\
D.~Lust, {\it Intersecting brane worlds: A path to the standard
model?}, Class.\ Quant.\ Grav.\ {\bf 21} (2004) S1399
[arXiv:hep-th/0401156];
\\
R.~D'Auria, S.~Ferrara and M.~Trigiante, {\it No--scale supergravity
from higher dimensions}, arXiv:hep-th/0409184;
\\
R.~Blumenhagen, M.~Cvetic, F.~Marchesano and G.~Shiu, {\it Chiral
D--brane models with frozen open string moduli}, arXiv:hep-th/0502095;
\\
K.~Behrndt, {\it Stabilization of moduli by fluxes}, AIP Conf.\ Proc.\
{\bf 743} (2005) 251 [arXiv:hep-th/0503129].
%
\bibitem{pol}
J.~Polchinski, {\em String theory} (Vol. 2: Superstring theory and
beyond), Cambridge University Press, 1998.
%
\bibitem{bc}
K.~Behrndt and M.~Cvetic, {\it General N = 1 supersymmetric fluxes in
massive type IIA string theory}, Nucl.\ Phys.\ B {\bf 708} (2005) 45
[arXiv:hep-th/0407263].
%
\bibitem{dkpz}
J.~P.~Derendinger, C.~Kounnas, P.~M.~Petropoulos and F.~Zwirner,
{\it Superpotentials in IIA compactifications with general fluxes},
Nucl.\ Phys.\ B {\bf 715} (2005) 211 [arXiv:hep-th/0411276].
%
\bibitem{lustsi}
D.~Lust and D.~Tsimpis, {\it Supersymmetric AdS(4) compactifications
of IIA supergravity}, JHEP {\bf 0502} (2005) 027
[arXiv:hep-th/0412250].
%
\bibitem{csu} 
M.~Cvetic, G.~Shiu and A.~M.~Uranga, {\it Three-family
supersymmetric standard like models from intersecting brane worlds},
Phys.\ Rev.\ Lett.\ {\bf 87} (2001) 201801 [arXiv:hep-th/0107143];
\\
M.~Cvetic, G.~Shiu and A.~M.~Uranga, {\it Chiral four-dimensional N =
1 supersymmetric type IIA orientifolds from intersecting D6--branes},
Nucl.\ Phys.\ B {\bf 615} (2001) 3 [arXiv:hep-th/0107166];
\\
R.~Blumenhagen, M.~Cvetic, F.~Marchesano and G.~Shiu, {\it Chiral
D--brane models with frozen open string moduli}, arXiv:hep-th/0502095.
%
\bibitem{romans}
L.~J.~Romans, {\it Massive N=2A Supergravity In Ten-Dimensions},
Phys.\ Lett.\ B {\bf 169} (1986) 374.
%
\bibitem{ss}
J.~Scherk and J.~H.~Schwarz, {\it How To Get Masses From Extra
Dimensions}, Nucl.\ Phys.\ B {\bf 153} (1979) 61.
%
\bibitem{km}
N.~Kaloper and R.~C.~Myers, {\it The O(dd) story of massive
supergravity}, JHEP {\bf 9905} (1999) 010 [arXiv:hep-th/9901045].
%
\bibitem{dagfer} 
G.~Dall'Agata and S.~Ferrara, {\it Gauged supergravity algebras from
twisted tori compactifications with fluxes}, arXiv:hep-th/0502066v2.
%
\bibitem{bkorv}
E.~Bergshoeff, R.~Kallosh, T.~Ortin, D.~Roest and A.~Van Proeyen, {\it
New formulations of D = 10 supersymmetry and D8 - O8 domain walls},
Class.\ Quant.\ Grav.\ {\bf 18}, 3359 (2001) [arXiv:hep-th/0103233].
%
\bibitem{gh}
S.~Gukov and M.~Haack, {\it IIA string theory on Calabi-Yau fourfolds
with background fluxes}, Nucl.\ Phys.\ B {\bf 639} (2002) 95
[arXiv:hep-th/0203267].
%
\bibitem{generalwIIA1}
G.~L.~Cardoso, G.~Curio, G.~Dall'Agata, D.~Lust, P.~Manousselis and
G.~Zoupanos, {\it Non-Kaehler string backgrounds and their five
torsion classes}, Nucl.\ Phys.\ B {\bf 652} (2003) 5
[arXiv:hep-th/0211118];
\\
G.~L.~Cardoso, G.~Curio, G.~Dall'Agata and D.~Lust, {\it BPS action
and superpotential for heterotic string compactifications with
fluxes}, JHEP {\bf 0310} (2003) 004 [arXiv:hep-th/0306088];
\\
G.~L.~Cardoso, G.~Curio, G.~Dall'Agata and D.~Lust, {\it Heterotic
string theory on non-Kaehler manifolds with H-flux and gaugino
condensate}, Fortsch.\ Phys.\ {\bf 52} (2004) 483
[arXiv:hep-th/0310021];
\\
K.~Becker, M.~Becker, K.~Dasgupta and S.~Prokushkin, {\it Properties
of heterotic vacua from superpotentials}, Nucl.\ Phys.\ B {\bf 666}
(2003) 144 [arXiv:hep-th/0304001];
%
\bibitem{generalwIIA2}
S.~Gukov, {\it Solitons, superpotentials and calibrations}, Nucl.\
Phys.\ B {\bf 574} (2000) 169 [arXiv:hep-th/9911011];
%
\\
S.~Gurrieri, J.~Louis, A.~Micu and D.~Waldram, {\it Mirror symmetry in
generalized Calabi-Yau compactifications}, Nucl.\ Phys.\ B {\bf 654}
(2003) 61 [arXiv:hep-th/0211102];
\\
M.~Grana, R.~Minasian, M.~Petrini and A.~Tomasiello, {\it
Supersymmetric backgrounds from generalized Calabi-Yau manifolds},
JHEP {\bf 0408} (2004) 046 [arXiv:hep-th/0406137];
\\
M.~Grana, R.~Minasian, A.~Tomasiello and M.~Petrini, {\it
Supersymmetric backgrounds from generalized Calabi-Yau manifolds}
Comptes Rendus Physique {\bf 5} (2004) 979;
\\
T.~W.~Grimm and J.~Louis, {\it The effective action of type IIA
Calabi-Yau orientifolds}, arXiv:hep-th/0412277.
%
\bibitem{gkp}
S.~B.~Giddings, S.~Kachru and J.~Polchinski, {\it Hierarchies from
fluxes in string compactifications}, Phys.\ Rev.\ D {\bf 66} (2002)
106006 [arXiv:hep-th/0105097];
\\
S.~Kachru, M.~B.~Schulz and S.~Trivedi, {\it Moduli stabilization from
fluxes in a simple IIB orientifold}, JHEP {\bf 0310} (2003) 007
[arXiv:hep-th/0201028];
\\
A.~R.~Frey and J.~Polchinski, {\it N = 3 warped compactifications},
Phys.\ Rev.\ D {\bf 65} (2002) 126009 [arXiv:hep-th/0201029].
%
\bibitem{eg}
K.~Behrndt, M.~Cvetic and P.~Gao, {\it General type IIB fluxes with
SU(3) structures}, arXiv:hep-th/0502154.
%
\bibitem{gvw}
S.~Gukov, C.~Vafa and E.~Witten, {\it CFT's from Calabi-Yau
four-folds}, Nucl.\ Phys.\ B {\bf 584} (2000) 69 [Erratum-ibid.\ B
{\bf 608} (2001) 477] [arXiv:hep-th/9906070].
%
\bibitem{stads4}
P.~Breitenlohner and D.~Z.~Freedman, {\it Positive Energy In Anti-De
Sitter Backgrounds And Gauged Extended Supergravity}, Phys.\ Lett.\ B
{\bf 115} (1982) 197;
\\
P.~Breitenlohner and D.~Z.~Freedman, {\it Stability In Gauged Extended
Supergravity}, Annals Phys.\ {\bf 144} (1982) 249;
\\
P.~K.~Townsend, {\it Positive Energy And The Scalar Potential In
Higher Dimensional (Super)Gravity Theories}, Phys.\ Lett.\ B {\bf 148}
(1984) 55;
\\
L.~Mezincescu and P.~K.~Townsend, {\it Stability At A Local Maximum In
Higher Dimensional Anti-De Sitter Space And Applications To
Supergravity}, Annals Phys.\ {\bf 160} (1985) 406.
%
\bibitem{magn}  
C.~Bachas, {\em A Way to break supersymmetry}, arXiv:hep-th/9503030;
\\ 
R.~Blumenhagen, L.~Goerlich, B.~Kors and D.~Lust, {\it Noncommutative
compactifications of type I strings on tori with magnetic background
flux}, JHEP {\bf 0010} (2000) 006 [arXiv:hep-th/0007024];
\\ 
C.~Angelantonj, I.~Antoniadis, E.~Dudas and A.~Sagnotti, {\it Type--I
strings on magnetised orbifolds and brane transmutation}, Phys.\
Lett.\ B {\bf 489} (2000) 223 [arXiv:hep-th/0007090];
\\
R.~Blumenhagen, D.~Lust and T.~R.~Taylor, {\it Moduli stabilization in
chiral type IIB orientifold models with fluxes}, Nucl.\ Phys.\ B {\bf
663} (2003) 319 [arXiv:hep-th/0303016];
\\
J.~F.~G.~Cascales and A.~M.~Uranga, {\it Chiral 4d N = 1 string vacua
with D--branes and NSNS and RR fluxes}, JHEP {\bf 0305} (2003) 011
[arXiv:hep-th/0303024];
\\
I.~Antoniadis and T.~Maillard, {\it Moduli stabilization from magnetic
fluxes in type I string theory}, arXiv:hep-th/0412008;
\\
A.~Font and L.~E.~Ibanez, {\it SUSY-breaking soft terms in a MSSM
magnetized D7-brane model}, arXiv:hep-th/0412150;
\\
M.~Bianchi and E.~Trevigne, {\it The open story of the magnetic
fluxes}, arXiv:hep-th/0502147.
%
\bibitem{soft}  
M.~Grana, {\it MSSM parameters from supergravity backgrounds}, Phys.\
Rev.\ D {\bf 67} (2003) 066006 [arXiv:hep-th/0209200];
\\
P.~G.~Camara, L.~E.~Ibanez and A.~M.~Uranga, {\it Flux-induced
SUSY-breaking soft terms}, Nucl.\ Phys.\ B {\bf 689} (2004) 195
[arXiv:hep-th/0311241];
\\
M.~Grana, T.~W.~Grimm, H.~Jockers and J.~Louis, {\it Soft supersymmetry
breaking in Calabi-Yau orientifolds with D--branes and fluxes}, Nucl.\
Phys.\ B {\bf 690} (2004) 21 [arXiv:hep-th/0312232];
\\
D.~Lust, S.~Reffert and S.~Stieberger, {\it Flux-induced soft
supersymmetry breaking in chiral type IIb orientifolds with
D3/D7-branes}, Nucl.\ Phys.\ B {\bf 706} (2005) 3
[arXiv:hep-th/0406092];
\\
P.~G.~Camara, L.~E.~Ibanez and A.~M.~Uranga, {\it Flux-induced
SUSY-breaking soft terms on D7-D3 brane systems}, Nucl.\ Phys.\ B {\bf
708} (2005) 268 [arXiv:hep-th/0408036];
\\
D.~Lust, S.~Reffert and S.~Stieberger, {\it MSSM with soft SUSY
breaking terms from D7-branes with fluxes}, arXiv:hep-th/0410074;
\\
D.~Lust, P.~Mayr, S.~Reffert and S.~Stieberger, {\it F-theory flux,
destabilization of orientifolds and soft terms on D7-branes},
arXiv:hep-th/0501139.
%
\end{thebibliography}
\end{document}